\documentclass[superscriptaddress,amsmath,amssymb,aps,prl,twocolumn]{revtex4-1}

\usepackage{times}
\usepackage{graphicx}
\usepackage{dcolumn}

\begin{document}

\title{Highly tunable elastic dielectric metasurface lenses}

\author{Seyedeh Mahsa Kamali}
\affiliation{T. J. Watson Laboratory of Applied Physics and Kavli Nanoscience Institute, California Institute of Technology, 1200 E California Blvd., Pasadena, CA 91125, USA}
\author{Ehsan Arbabi}
\affiliation{T. J. Watson Laboratory of Applied Physics and Kavli Nanoscience Institute, California Institute of Technology, 1200 E California Blvd., Pasadena, CA 91125, USA}
\author{Amir Arbabi}
\affiliation{T. J. Watson Laboratory of Applied Physics and Kavli Nanoscience Institute, California Institute of Technology, 1200 E California Blvd., Pasadena, CA 91125, USA}
\author{Yu Horie}
\affiliation{T. J. Watson Laboratory of Applied Physics and Kavli Nanoscience Institute, California Institute of Technology, 1200 E California Blvd., Pasadena, CA 91125, USA}
\author{Andrei Faraon}
\email{Corresponding author: A.F: faraon@caltech.edu}
\affiliation{T. J. Watson Laboratory of Applied Physics and Kavli Nanoscience Institute, California Institute of Technology, 1200 E California Blvd., Pasadena, CA 91125, USA}

\begin{abstract}
Dielectric metasurfaces are two-dimensional structures composed of nano-scatterers that manipulate phase and polarization of optical waves with subwavelength spatial resolution, enabling ultra-thin components for free-space optics. While high performance devices with various functionalities, including some that are difficult to achieve using conventional optical setups have been shown, most demonstrated components have a fixed functionality. Here we demonstrate highly tunable metasurface devices based on subwavelength thick
silicon nano-posts encapsulated in a thin transparent elastic polymer. As proof of concept, we demonstrate a metasurface microlens operating at 915 nm, with focal distance tuning from 600 $\mu$m to 1400 $\mu$m through radial strain, while maintaining a diffraction limited focus and a focusing efficiency above 50$\%$. The demonstrated
tunable metasurface concept is highly versatile for developing ultra-slim, multi-functional and tunable optical devices with widespread applications ranging from consumer electronics to medical devices and optical communications.
\end{abstract}

\maketitle
Metasurfaces are composed of a large number of discrete nano-scatterers (meta-atoms) that locally modify phase and polarization of light with subwavelength spatial resolution. The meta-atoms can be defined lithographically, thus providing a way to mass-produce thin optical elements \cite{Kildishev2013Science,Yu2014NatMater,Lin2014Science,Arbabi2015NatNano} that could directly replace traditional bulk optical components or provide novel functionalities \cite{Arbabi2015NatNano,Kamali2015}. The two dimensional nature and the subwavelength thickness of metasurfaces make them suitable for tunable and reconfigurable optical elements. Some efforts have recently been focused on developing tunable and reconfigurable metasurfaces using different stimuli for tuning the meta-atoms. Examples include frequency response tuning using substrate deformation \cite{Zhu2015Optica,Gutruf2016ACSNano}, refractive index tuning via thermo-optic effects \cite{Donner2015ACSPhotonics,Sautter2015ACSNano}, phase change materials~\cite{Chen2015SciRep, Wang2016NatPhoton}, and electrically driven carrier accumulation \cite{Yao2014NanoLet, Huang2015}.

Stretchable substrates have been used to demonstrate tunable diffractive and plasmonic metasurface components \cite{Simonov2005OptLett,Li2015IEEE,Ee2016NanoLett}, but they have exhibited low tunability, poor efficiency, polarization dependent operation, or significant optical aberrations. Here we present mechanically tunable dielectric metasurfaces based on elastic substrates, simultaneously providing a high tuning range, polarization independence, high efficiency, and diffraction limited performance. As a proof of principle, we experimentally demonstrate an aspherical microlens with over 130$\%$ focal distance tuning (from 600 $\mu$m to 1400 $\mu$m) while keeping high efficiency and diffraction limited focusing.

\begin{figure}[htp]
\centering
\includegraphics{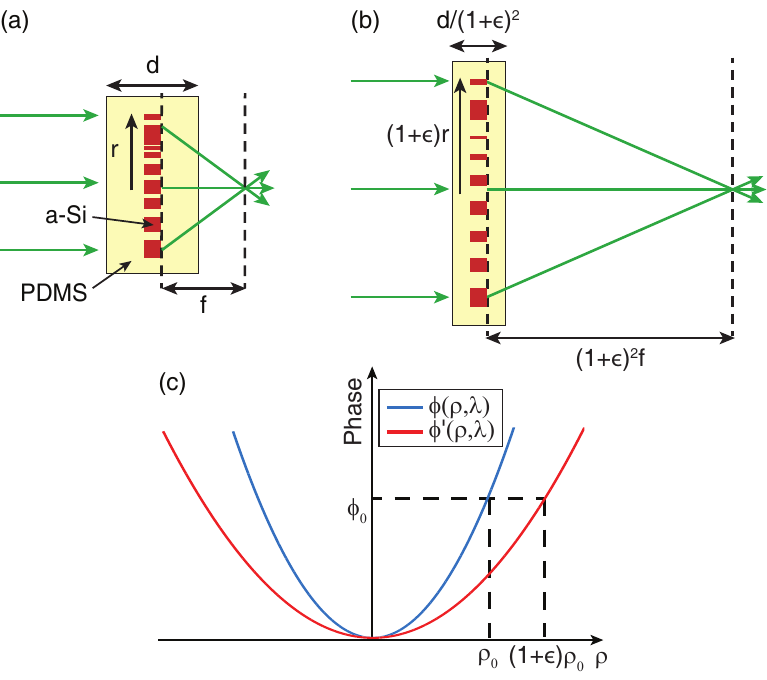}
\caption{Principle of tunable elastic metasurface lenses. (a) A side view schematic illustration of a dielectric metasurface microlens with focal distance of $f$ encapsulated in a low index elastic membrane. (b) By stretching the metasurface microlens with a stretch ratio of $1+\epsilon$, its focal distance changes by $(1+\epsilon)^2$, providing a large tunability. The membrane thickness decreases according to its Poisson ratio ($\nu$), considered to be 0.5 here. (c) Phase of the metasurface microlens before (solid blue curve) and after (solid red curve) stretching. a-Si: amorphous silicon, PDMS: Polydimethylsiloxane.}
\label{fig:1_concept}
\end{figure}

Figure \ref{fig:1_concept}(a) shows a schematic of a metasurface microlens encapsulated in an elastic substrate with radius $r$ and focal distance $f$. The phase profile of the lens has the following form, and is drawn in Fig. \ref{fig:1_concept}(c) (solid blue curve):
\begin{equation}
\phi(\rho,\lambda) = \frac{2\pi}{\lambda}(\sqrt{\rho^2 + f^2}-f) \label{eq:unstretched_phase}\\
\end{equation}

where $\rho$ is the distance to the center of the lens. Equation (\ref{eq:unstretched_phase}) in the paraxial approximation ($\rho\ll f$) reduces to 

\begin{equation}
\phi(\rho,\lambda)\approx \frac{\pi \rho^2}{\lambda f} \label{eq:unstretched_phase_paraxial}\\
\end{equation}

By uniformly stretching the substrate with a stretch ratio of $1+\epsilon$, and assuming that the local phase transformation does not depend on the substrate deformation, the phase initially applied at radius $\rho$ is now applied at radius $\rho(1+\epsilon)$; therefore,  the under strain phase profile becomes $\phi'(\rho,\lambda) = \pi\rho^2/(\lambda(1+\epsilon)^2f)$ (shown in Fig. \ref{fig:1_concept}(c), solid red curve). This indicates that by stretching the elastic metasurface microlens with stretching ratio of $1+\epsilon$ the focal length scales by a factor of $(1+\epsilon)^2$, as shown schematically in Fig. \ref{fig:1_concept}(b).

\begin{figure}[htp]
\centering
\includegraphics{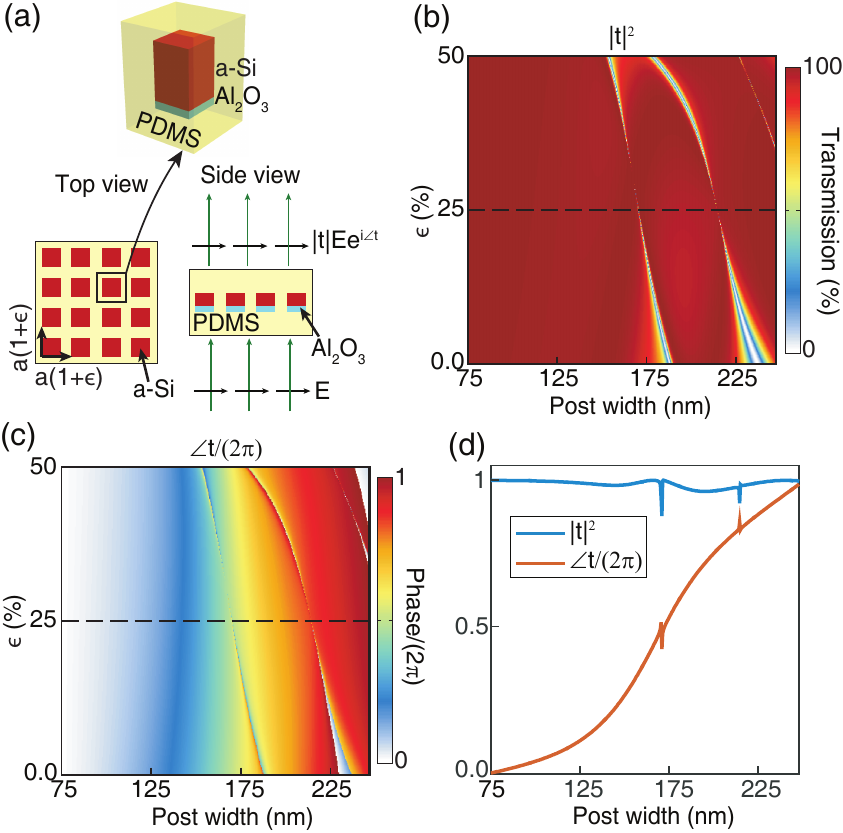}
\caption{Design procedure of tunable metasurfaces. (a) Schematic drawing of the top and side views of a uniform array of square cross-section nano-posts arranged in a square lattice and encapsulated in PDMS. The inset shows the building block of the array: an amorphous silicon nano-post on a thin layer of aluminum oxide. (b) Simulated intensity and (c) phase of the transmission coefficient for the array shown in (a) as a function of the nano-post width and the substrate strain. (d) Simulated intensity and phase of the transmission coefficient for $\epsilon=25\%$ (corresponding to the dashed lines shown in (b) and (c)) used to map the transmission values to the nano-post widths. The nano-posts are 690 nm tall, aluminum oxide layer is $\sim$100 nm thick, and the simulation wavelength is 915 nm. a-Si: amorphous silicon, PDMS: Polydimethylsiloxane.}
\label{fig:2_design}
\end{figure}

\begin{figure}[htp]
\centering
\includegraphics{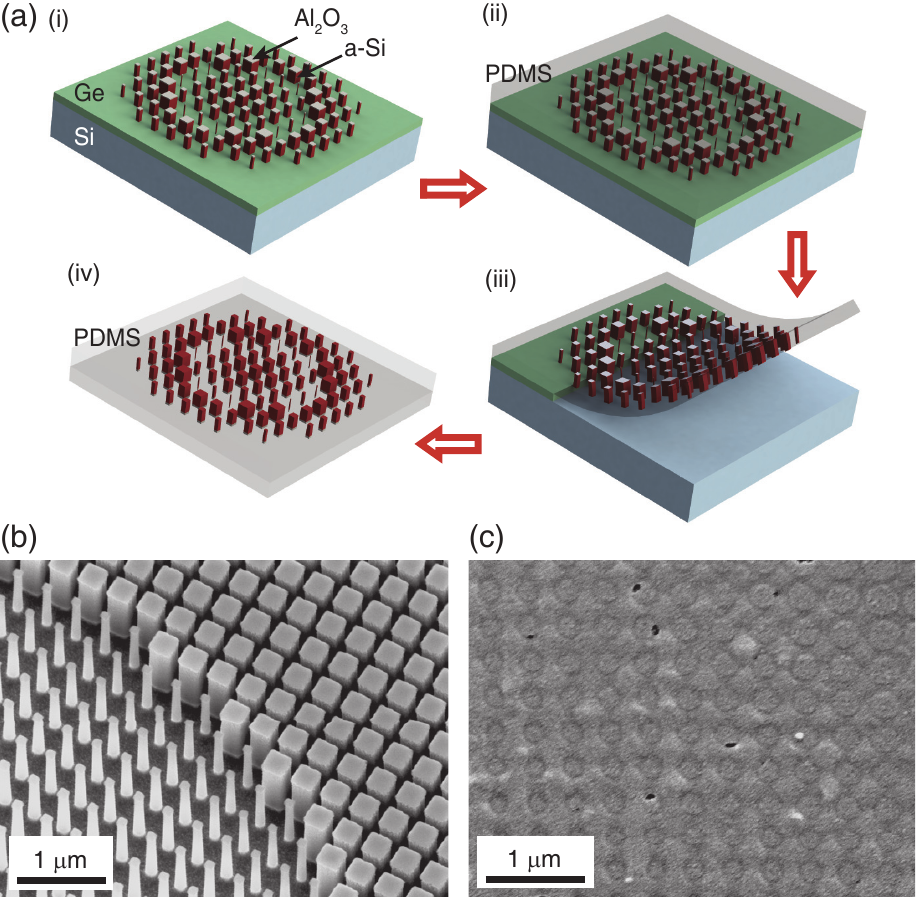}
\caption{Overview of the fabrication steps and images of the device at different steps. (a) Major steps involved in fabricating tunable metasurfaces: (i) Amorphous silicon nano-posts are patterned and dry etched using an aluminum oxide hard mask. The nano-posts rest on a germanium sacrificial layer on a silicon wafer. (ii) PDMS is spin coated on the metasurface structure. (iii) The sacrificial germanium layer is dissolved to release the nano-posts which are now embedded in the flexible and stretchable PDMS layer. (iv) A second PDMS layer is spin coated on the side containing the metasurface to provide a fully encapsulated microlens. (b) Scanning electron micrograph of the nano-posts before spin coating the first PDMS layer (step (i)). (c)  Scanning electron micrograph of the nano-posts embedded in PDMS (step (iii)), taken at a tilt angle of 30 degrees. To dissipate the electric charge accumulated during scanning electron microscopy, a $\sim$20-nm-thick gold layer was deposited on the sample prior to imaging. Small holes observed around the nano-posts are in the deposited gold layer. a-Si: amorphous silicon, PDMS: Polydimethylsiloxane.}
\label{fig:3_Fab}
\end{figure}

\begin{figure*}[htb]
\includegraphics{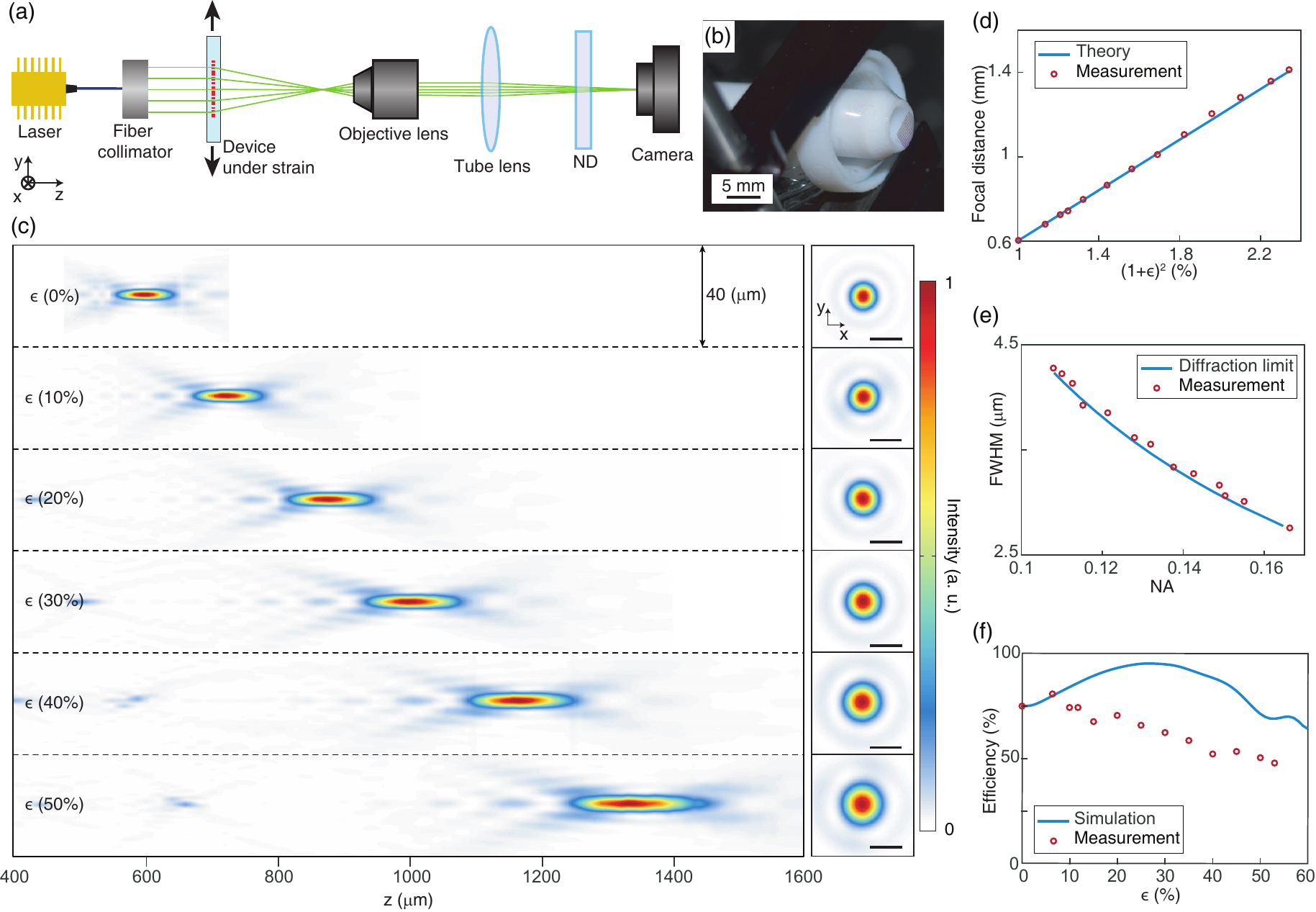}
\caption{Measurement results. (a) Schematic of the measurement setup, ND: neutral density filter. (b) An array of elastic metasurface microlenses clamped between two Teflon rings, under $\sim$30$\%$ strain. (c) Measured optical intensity profiles of a radially strained metasurface microlens ($\epsilon$ = 0$\%$ to 50$\%$) in the axial plane (left) and the focal plane (right).  Scale bars: 5 $\mu$m. (d) Measured and analytically predicted focal distances (i. e. $(1+\epsilon)^2f$) for different strain values versus square of the stretch ratio ($(1+\epsilon)^2$). Focal distance is tuned from 600 $\mu$m to more than 1400 $\mu$m. (e) Measured and diffraction limited  full width at half maximum (FWHM) spot size in the focal plane for different strain values as a function of the numerical aperture (NA) of the microlens. (f) Measured and simulated focusing efficiencies as a function of strain. Measurements and simulations are performed at the wavelength of 915 nm. See Supplement 1 for the measurement and simulation details.}
\label{fig:4_Measurement}
\end{figure*}

For implementation, we used a metasurface platform based on high contrast dielectric meta-atoms placed on a subwavelength periodic lattice in a low index medium. The building blocks of the metasurface are amorphous silicon square nano-posts on a thin layer of aluminum oxide encapsulated in Polydimethylsiloxane (PDMS) as a low index elastic membrane (Fig. \ref{fig:2_design}(a), inset).

A key characteristic of this platform is the weak optical coupling between the nano-posts, which simplifies the metasurface design by allowing local sampling of the phase profile using different widths for the nano-posts placed on the vertices of a square lattice. This weak coupling is due to the high index contrast between the nano-posts and their surrounding medium, and is manifested in the high localization of energy density inside the nano-posts \cite{Arbabi2015NatCommun}. An important consequence of the weak coupling is that the phase transformation mainly depends on the nano-posts width and not on the distance between them, leading to the same local phase shift almost independent of the stretch factors of the substrate. Figures \ref{fig:2_design}(b) and \ref{fig:2_design}(c) show the simulated transmittance and phase of the transmission coefficient for a periodic square lattice of encapsulated nano-posts in PDMS with strain values from 0$\%$ to 50$\%$. The nano-posts are assumed to be 690 nm tall, and the lattice constant at 0$\%$ strain is 380 nm. Nano-posts height must be chosen such that the whole 0 to 2$\pi$ phase range is covered at all strains of interest by changing the nano-posts width, while keeping high transmission values. The lattice constant should be selected such that the lattice is subwavelength and satisfies the Nyquist sampling criterion simultaneously for all strain values (Supplement 1 and Fig. S1). The simulation results are obtained assuming normal incident angle at the wavelength of 915 nm (see Supplement 1 for simulation details). The weak dependence of the transmission of the nano-post array to different strain values, which can be seen in Figs. \ref{fig:2_design}(b) and \ref{fig:2_design}(c), is another evidence for the weak coupling between the nano-posts.

Because the transmission coefficient is almost independent of the strain, we can design the metasurface at one specific strain. Figure \ref{fig:2_design}(d) shows the intensity and phase of the transmission coefficient at the middle strain value ($\epsilon$ = 25$\%$) as a function of the nano-post width, that is used for designing the tunable metasurface. Considering the desired phase profile $\phi(\rho)$ at 25$\%$ strain as given by Eq. (\ref{eq:unstretched_phase}), the corresponding nano-post width at each lattice site was found by minimizing the transmission error $\Delta T = |e^{i\phi} - |t|e^{i\angle t}|$, where $t$ is the complex transmission coefficient. Minimizing $\Delta T$ at each lattice site results in selecting the nano-post with the closest complex transmission value to the desired one ($\mathrm{e}^{i\phi}$) and automatically excludes the two high quality factor resonances observed in Fig. \ref{fig:2_design}(b) around 171 nm and 214 nm nano-post widths.

Using the proposed platform, a tunable metasurface microlens is designed to operate at the wavelength of 915 nm. The microlens has a diameter of 200 $\mu$m under no strain, and its focal distance changes from 600 $\mu$m to more than 1400 $\mu$m when the strain value varies from 0$\%$ to 53$\%$, exhibiting more than 130$\%$ tunability.

Figure \ref{fig:3_Fab}(a) schematically illustrates the key steps in fabricating a metasurface encapsulated in a thin elastic membrane. A germanium layer followed by an amorphous silicon layer are deposited on a silicon wafer. The pattern is defined using electron beam lithography and is transferred to the amorphous silicon layer by dry etching and using an aluminum oxide hard mask (Fig. \ref{fig:3_Fab}(a), (i)). A diluted thin layer of PDMS followed by a thicker layer are subsequently spin coated on the sample (Fig. \ref{fig:3_Fab}(a), (ii)). The PDMS layer containing the metasurface is released by dissolving the germanium layer in a diluted ammonia solution (Fig. \ref{fig:3_Fab}(a), (iii)). A second layer of PDMS is spin coated on the metasurface side of the sample to completely encapsulate the nano-posts in PDMS (Fig. \ref{fig:3_Fab}(a), (iv)). The PDMS forms a monolithic $\sim$100-$\mu$m-thick layer (see Supplement 1 for fabrication details). Encapsulation of nano-posts in PDMS is a crucial step in preserving the metasurface shape and minimizing defects when the device is highly strained (see Supplement 1 Fig. S2). A scanning electron micrograph of the nano-posts on germanium layer before spin coating the first PDMS layer is shown in Fig. \ref{fig:3_Fab}(b). The nano-post transfer process has a near unity yield in retaining almost all the nano-posts at their positions~\cite{Kamali2015}. Void-free filling of the gaps between the nano-posts was confirmed by inspecting nano-posts embedded in PDMS before spin coating the PDMS cladding (Fig. \ref{fig:3_Fab}(c)).

For characterization of the fabricated tunable metasurface microlens, a custom built microscope was used to image the transmitted light intensity at different distances from the metasurface (Fig. \ref{fig:4_Measurement}(a)). First, the sample was mounted on a flat glass substrate and was characterized in the relaxed mode, and then it was clamped between two Teflon rings. A radial force was applied by pushing a Teflon tube from the backside and stretching the metasurface. An array of microlenses mounted between the rings and under $\sim$30$\%$ strain is shown in Fig. \ref{fig:4_Measurement}(b). (see Supplement 1 for measurement details). Measured optical intensities in the axial plane (Fig. \ref{fig:4_Measurement}(c), left) and the focal plane (Fig. \ref{fig:4_Measurement}(c), right) at 6 different strain values (0$\%$ to 50$\%$) show a large focal distance tunability while keeping a nearly diffraction limited focus at all strains. The measured intensity profiles at different strain values are in good agreement with the simulated intensity profiles shown in Supplement 1 Fig. S3. Figure \ref{fig:4_Measurement}(d) shows a good agreement between the measured and the analytically predicted focal distances, which are plotted versus $(1+\epsilon)^2$. Measured full width at half maximum (FWHM) of the focal spots for different strains and their corresponding diffraction limited values are shown in Fig. \ref{fig:4_Measurement}(e) as a function of the numerical aperture (NA) of the microlens. The results show nearly diffraction limited operation of the microlens under strain values up to above 50$\%$. As expected,  NA decreases and focal spot enlarges as strain is increased.  The measured and simulated focusing efficiencies for different strains are shown in Fig. \ref{fig:4_Measurement}(f). The focusing efficiency is defined as the ratio of the optical power focused by the device to the incident power (see Supplement 1 for simulation and measurement details).

The measured focusing efficiency of $\sim$75$\%$ for the relaxed device demonstrates its high optical quality. The efficiency decreases gradually with increasing the strain; however, it remains above 50$\%$ for strain values up to 50$\%$. The simulated focusing efficiency peaks close to the 25$\%$ strain, for which the device was designed. At small strains, the measured focusing efficiency agrees well with its simulated value, but the measured efficiency is lower at large strain values. We attribute the lower measured efficiency to possible mechanical deformations and misalignments of the nano-posts under strain, and the non-uniformity of the strain across the microlens.

The reliability of the tuning process was tested by measuring the focal spot and focusing efficiency of the tunable metasurface microlens after multiple straining cycles. No changes in the focusing efficiency and focal spot shape of the microlens was observed after more than 10 cycles of stretching and releasing the device (see Supplement 1 Fig. S4).

In conclusion, we demonstrated highly tunable dielectric metasurfaces based on elastic substrates. As proof of concept, a spherical-aberration-free microlens with more than 130$\%$ focal length tunability was demonstrated. The proposed platform can be applied to other devices based on metasurfaces thus adding tunability over a thin layer without increasing the complexity of the system. Tunable metasurfaces can also be fabricated on high speed electrically tunable elastomers in order to decrease their response time to less than a millisecond \cite{Pelrine2000Science}. Moreover, integration of this platform with flexible and wearable electronics \cite{Rogers2010Science} can lead to versatile tunable optoelectronic technologies.

\section*{Acknowledgement}
This work was supported by the DOE ``Light-Material Interactions in Energy Conversion" Energy Frontier Research Center funded by the US Department of Energy, Office of Science, Office of Basic Energy Sciences under Award no. DE-SC0001293. E.A., A.A., and Y.H. were supported by Samsung Electronics. The device nanofabrication was performed at the Kavli Nanoscience Institute at Caltech.

\clearpage

\newcommand{\beginsupplement}{%
        \setcounter{table}{0}
        \renewcommand{\thetable}{S\arabic{table}}%
        \setcounter{figure}{0}
        \renewcommand{\thefigure}{S\arabic{figure}}%
     }

\onecolumngrid

      \beginsupplement

\section{Supplementary Information for ``Highly tunable elastic dielectric metasurface lenses"}

\section{Methods}

\noindent\textbf{Design procedure}

To find the transmittance and phase values (Fig. 2(b-d)), a periodic array of square nano-posts on a square lattice was simulated at 915 nm with a normally incident plane wave using rigorous coupled wave analysis (RCWA)~\cite{Liu2012CompPhys}. Refractive indices of 3.56 and 1.41 were used for a-Si and PDMS. The lattice constant was chosen to be 380 nm at 0$\%$ strain and linearly scaled with the stretch ratio. It was chosen such that the array remains non-diffractive with enough sampling unit cells for reconstructing the wavefront at all the strain values of interest (Supplement 1 section 2). The metasurface microlens was designed for the middle strain (25$\%$ strain), for which the lattice constant is 475 nm. The lattice constant was then scaled down to 380 nm for device fabrication.

\noindent\textbf{Simulation}

The intensity distributions shown in Fig. \ref{fig:S3_simulation} were found by modeling the microlens as a phase mask. The transmission coefficient of the phase mask was calculated through interpolation of the complex transmission coefficients of the nano-posts. The effect of the strain was considered in both the position and the transmission coefficient of the nano-posts. A plane wave was used to illuminate the phase mask. The fields after the phase mask were propagated through the top PDMS layer ($\sim$50 $\mu$m thick at zero strain) and air to the focal plane and beyond using plane wave expansion technique. For efficiency calculations, a Gaussian beam with more than 99$\%$ of its power inside the device was used. The Gaussian beam radius was linearly scaled with the stretch ratio. Intensity profiles in the focal plane for different strain values were found using the same plane wave expansion technique. The focusing efficiencies were calculated by dividing the power passing through a disk around the focal point to the total incident power. The diameter of the disk for each strain value was set to be $\sim$3 times the analytical FWHM. The devices were also simulated using the finite difference time domain method \cite{Oskooi2010Computer} and the intensity distributions and the focusing efficiencies were in good agreements with the described simulation method.

\noindent\textbf{Sample fabrication}

A germanium sacrificial layer ($\sim$300 nm) was evaporated on a silicon wafer, followed by a 690-nm-thick hydrogenated PECVD (plasma enhanced chemical vapor deposition) a-Si layer (5$\%$ mixture of silane in argon at $200\,^{\circ}{\rm C}$). The refractive index of the a-Si layer was found to be 3.56 at the wavelength of 915 nm, using variable angle spectroscopic ellipsometry. A Vistec EBPG5000+ e-beam lithography system was used to define the pattern in ZEP-520A positive resist ($\sim$300 nm, spin coated at 5000 rpm for 1 min). A resist developer (ZED-N50 from Zeon Chemicals) was used to develop the pattern for 3 minutes. A $\sim$100-nm-thick aluminum oxide layer was deposited on the sample by e-beam evaporation. The pattern was then transferred into aluminum oxide by lifting off the resist. The patterned aluminum oxide hard mask was used for dry etching the a-Si layer in a mixture of $\mathrm{SF_6}$ and $\mathrm{C_4F_8}$ plasma. The PDMS (10:1 mixing ratio of Sylgard 184 base and curing agent) was diluted in toluene in a 2:3 weight ratio as a thinner. The diluted PDMS mixture was spin coated (at 3000 rpm for 1 min) on the fabricated metasurface to fill the gaps between the nano-posts and to form a thin PDMS film. The sample was then degassed and cured at $80\,^{\circ}{\rm C}$ for more than 30 mins. The second layer of PDMS without a thinner ($\sim$50 $\mu$m, spin coated at 1000 rpm for 1 min) was likewise degassed and cured at $80\,^{\circ}{\rm C}$ for more than 1 hr. The sample was then immersed in a 1:1:30 mixture of ammonium hydroxide, hydrogen peroxide, and DI water at room temperature to remove the sacrificial germanium layer and release the embedded nano-posts in the PDMS substrate ($\sim$ one day). Another layer of PDMS without a thinner was then spin coated on the microlens side of the sample (at 1000 rpm for 1 min) to fully encapsulate the nano-posts. The sample was again degassed and cured at $80\,^{\circ}{\rm C}$ for more than 1 hr. The total PDMS thickness was $\sim$100 $\mu$m. To compensate for systematic fabrication errors, an array of devices with all the nano-post widths biased uniformly in steps of 3 nm was fabricated (Fig. 4b).

\noindent\textbf{Measurement procedure}

The device was measured using the setup shown schematically in Fig. 4a. A 915 nm fiber coupled semiconductor laser was used for illumination and a fiber collimation package (Thorlabs F220APC-780) used to collimate the incident beam. A 50X objective lens (Olympus LMPlanFL N, NA=0.5) and a tube lens (Thorlabs LB1945-B) with a focal distance of 20 cm were used to image intensity at different planes to a camera (CoolSNAP K4 from Photometrics). To adjust the light intensity and decrease the background noise captured by the camera, neutral density (ND) filters (Thorlabs ND filters, B coated) were used. A calibration sample with known feature sizes was also imaged with the setup to find the overall magnification.  The sample was first mounted on a glass substrate, for characterization under no strain. Then for the measurements under strain it was manually clamped between two machined Teflon rings, such that the microlens of interest was placed near the center of the rings. Then the clamped sample mounted on a translation stage was pushed toward a machined Teflon tube, such that the microlens of interest was stretched radially. 

To measure the focusing efficiencies under a specific strain, an additional lens with a focal length of 10 cm (Thorlabs LB1676-B) was used to partially focus the collimated beam. The beam radius was changed by adjusting the relative distance between the lens and the device under the test, such that more than 99\% of the beam power falls inside the device under the test. A pinhole with a diameter $\sim$3 times the measured FWHM was placed in the focal plane of the microlens to only let the light inside the pinhole pass through. The pinhole was fabricated by evaporating a thick layer of chrome on a fused silica substrate, and defining holes in it by wet etching. A power meter (Thorlabs PM100D) with a photodetector (Thorlabs S122C) was used to measure efficiencies at 915 nm. The focusing efficiency was calculated as the ratio of the power in focus (measured optical power after the pinhole) to the incident power (measured power before the sample). The focusing efficiency at 15$\%$ strain was measured in this manner. Focusing efficiencies at other strains were calculated relative to the focusing efficiency at 15$\%$ strain in the following manner: first, light intensity captured with the camera in the plane of focus was integrated inside a circle with a diameter $\sim$3 times of the measured FWHM at each strain value including the 15$\%$ strain. Then, the integrated power for each strain was divided by the integrated power at 15$\%$ strain. Moreover, the ratio of the input power at 15$\%$ strain to the input power at other strains was calculated (the input power of the beam hitting the device increases as the device area increases). The focusing efficiency at other strains was then found by multiplying these two normalization factors by the directly measured efficiency at 15$\%$ strain. The measurement setup used for the efficiency characterization is shown in Fig. \ref{fig:S5_Setup}.

\section{Sampling frequency of the phase profile}

\begin{figure*}[htp]
\centering
\includegraphics{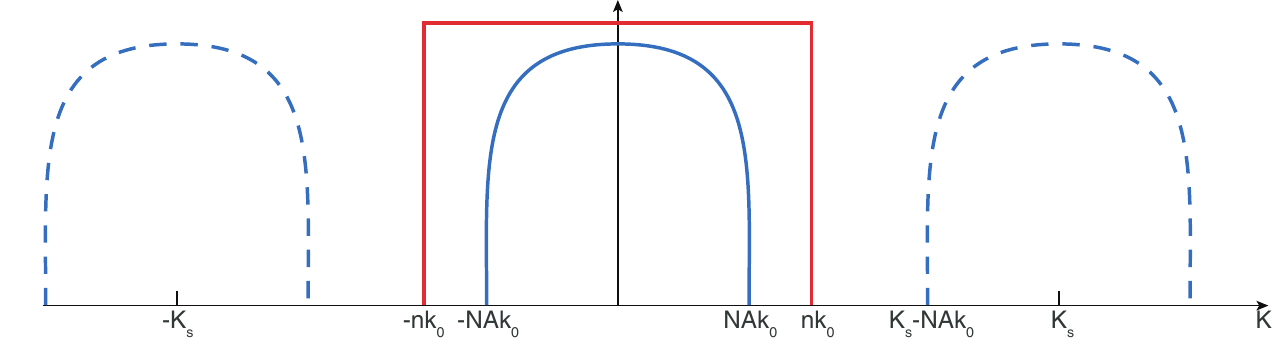}
\caption{Sampling frequency of the phase profile for perfect reconstruction of the wavefront. The flat microlens locally varying transmission coefficient spectrum can be considered as a band-limited signal with $2\mathrm{NA}k_0$ bandwidth (solid blue curve). By sampling the transmission coefficient with sampling frequency of $K_s$, displaced copies of the band-limited signal are added to the spectrum (dashed blue curves). In order to avoid undesirable diffractions, the free space low pass filter (solid red curve) should only filter the zeroth order diffraction (solid blue curve).}
\label{fig:S1_SamplingFreq}
\end{figure*}

The lattice constant should be chosen such that the lattice remains non-diffractive and satisfies the Nyquist sampling  criterion. From a signal processing point of view, the locally varying transmission coefficient of a flat microlens can be considered as a spatially band-limited signal with a $2\mathrm{NA}k_0$ bandwidth (ignoring the effect of the edges), where $\mathrm{NA}$ is the microlens numerical aperture, and $k_0$ is the vacuum wavenumber. A hypothetical one dimensional band-limited spectrum is depicted in Fig. \ref{fig:S1_SamplingFreq} (solid blue curve). By sampling the microlens phase profile with sampling frequency of $K_s$, the images (dashed blue curves in Fig. \ref{fig:S1_SamplingFreq}) are added to the spectrum. Therefore, for the perfect reconstruction of the microlens' transmission coefficient the Nyquist criterion should be satisfied: $K_s>2\mathrm{NA}k_0$.
On the other hand the lattice should remain subwavelength; the higher order diffractions (dashed blue curves in Fig. \ref{fig:S1_SamplingFreq}) should remain non-propagating. Propagation in free space can be considered as a low pass filter with $2nk_0$ bandwidth (solid red curve in Fig. \ref{fig:S1_SamplingFreq}), where $n$ is the the refractive index of the surrounding medium. Therefore, in order to have perfect reconstruction of phase and non-propagating higher order diffractions, the following relation should be satisfied:
\begin{equation}
K_s>nk_0+ \mathrm{NA} k_0
\label{eq:Sampling_Freq}
\end{equation}

Note that the sampling frequency ($K_s$) is a reciprocal lattice vector. For the square lattice $K_s=2\pi/\Lambda$, where $\Lambda$ is the lattice constant. Therefore Eq. (\ref{eq:Sampling_Freq}) would be simplified as follows:

\begin{equation}
\Lambda < \frac{\lambda}{n+\mathrm{NA}}
 \label{eq:Sampling_Freq_cubic_lattice}\\
\end{equation}

Where $\lambda$ is the free space wavelength. Note that the maximum value of numerical aperture is $\mathrm{NA_{max}}=n$, which simplifies Eq. (\ref{eq:Sampling_Freq_cubic_lattice}) to $\Lambda < \lambda/(2n)$. For designing tunable microlenses, Eq. (\ref{eq:Sampling_Freq_cubic_lattice}) should be satisfied for all the strains of interest, and $\Lambda = (1+\epsilon)a$, where $a$ is the lattice constant under strain. For the parameters used in the main text, the  unstrained lattice constant should be smaller than 401 nm in order to have tunable microlens up to 50$\%$ strains. The unstrained lattice constant was chosen to be 380 nm.

\begin{figure*}[htp]
\centering
\includegraphics[width=0.6\columnwidth]{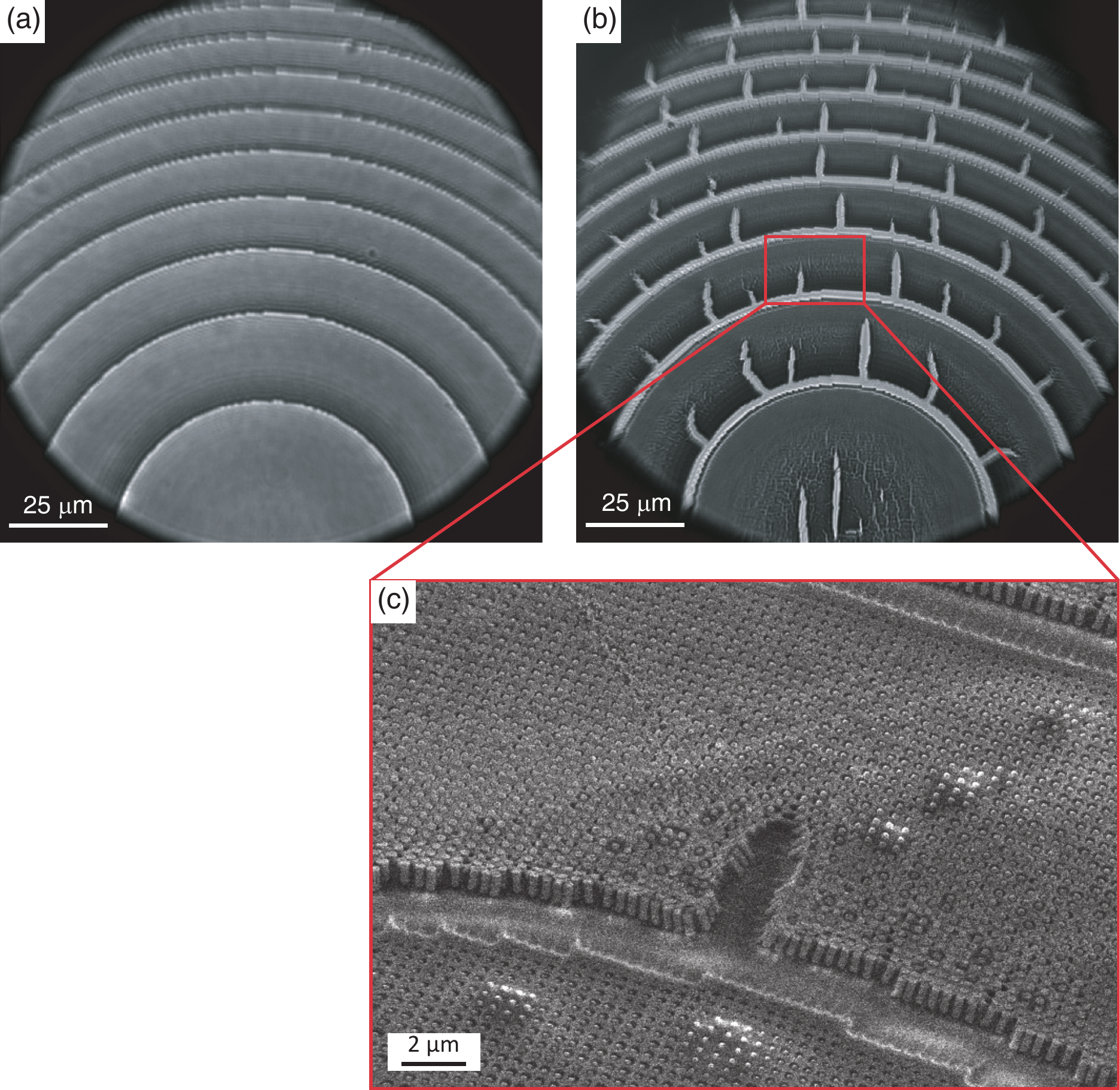}
\caption{Importance of PDMS cladding in the performance of the elastic metasurface under high strains. Optical images of the nano-posts in PDMS with (a) and without the PDMS claddings (b) under $\sim$50$\%$ radial strain. The images are taken using the same measurement setup shown in Fig. 4(a) under green laser illumination. Elastic metasurface without the PDMS cladding stretches non-uniformly, and some cracks are formed at the borders of the small and large nano-posts starting at $\sim$25$\%$ strain. By increasing the strain, these cracks spread in the elastic metasurface and some of the nano-posts stick out of the PDMS. (c) Scanning electron micrograph of the nano-posts without the PDMS cladding under $\sim$50$\%$ radial strain, taken at a tilt angle of 30 degrees. a $\sim$10-nm-thick gold layer was deposited on the sample to dissipate charge accumulation during the scanning electron imaging. The metasurface microlens presented in the main manuscript has PDMS cladding, and its nano-posts are completely encapsulated inside a thin PDMS layer. In this manner, the cracks do not show up between the nano-posts even at very high strains (as shown in (a)).}
\label{fig:S2_Encapsulated}
\end{figure*}

\begin{figure*}[htp]
\centering
\includegraphics[width=\columnwidth]{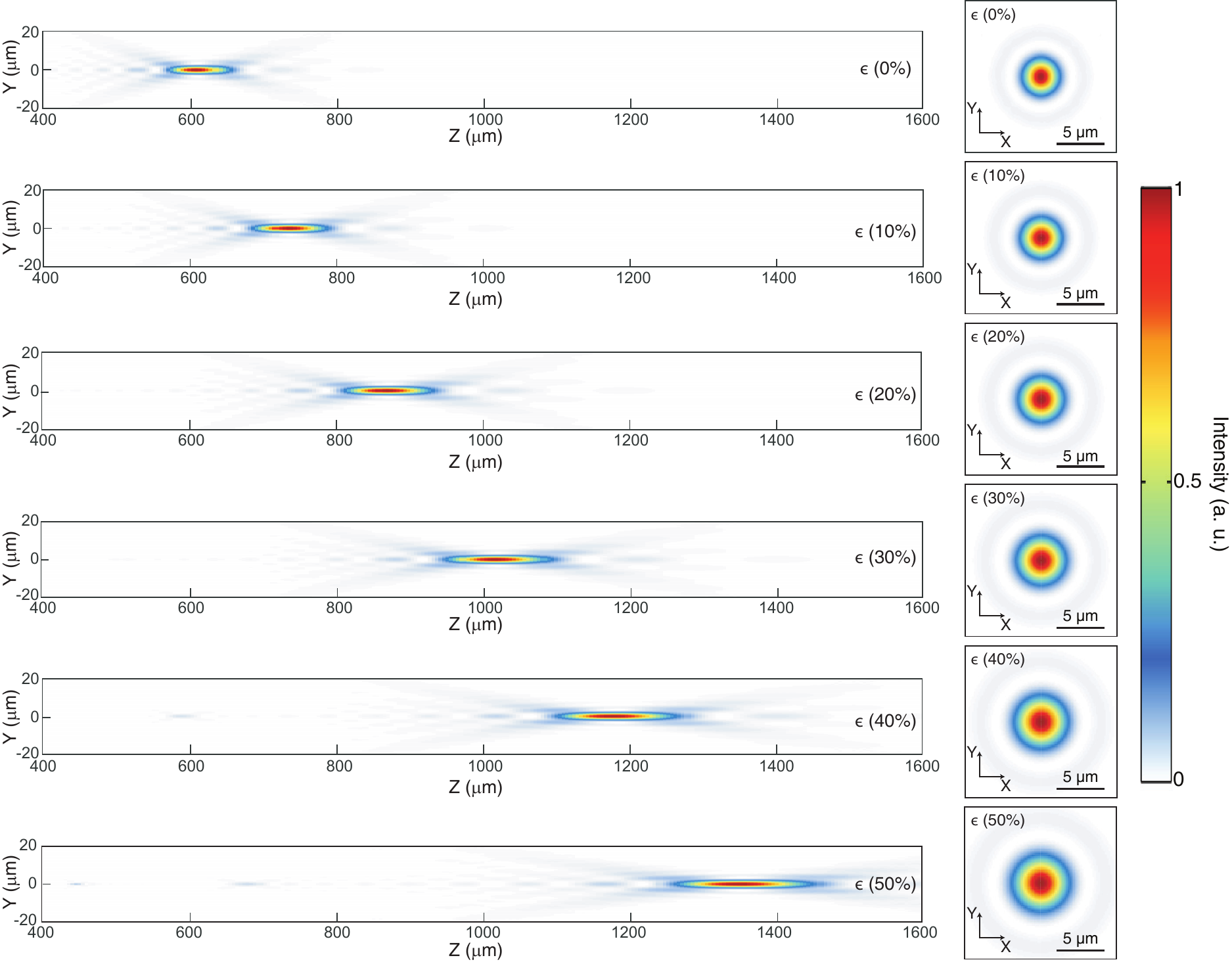}
\caption{Simulation results of the tunable microlens. Simulated intensity profiles for different strain values ($\epsilon$ = 0$\%$ to 50$\%$) are shown in the axial plane (left) and in the focal plane (right).}
\label{fig:S3_simulation}
\end{figure*}

\begin{figure*}[htp]
\centering
\includegraphics[width=0.8\columnwidth]{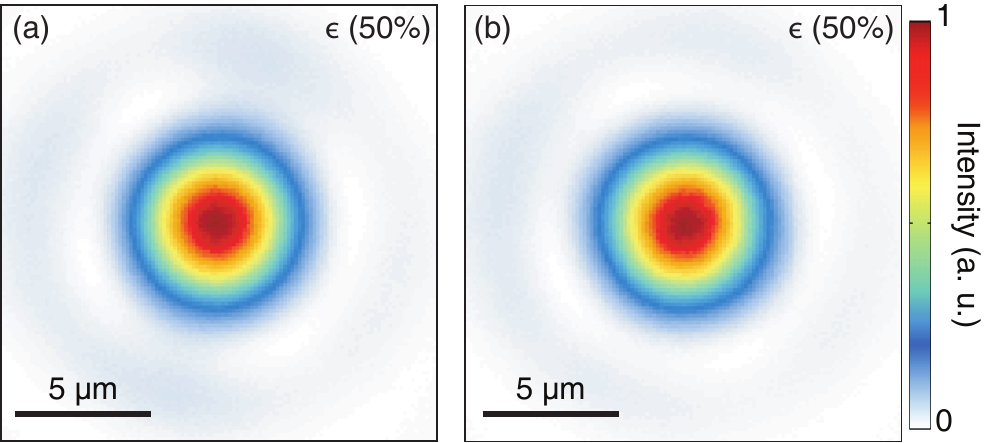}
\caption{Reliability measurement of the microlens under strain. Measured optical intensity profile of the the tunable microlens in the plane of focus under 50$\%$ strain after one (a), and more than ten times of stretching and releasing the device (b).}
\label{fig:S4_reliability}
\end{figure*}

\begin{figure*}[htp]
\centering
\includegraphics{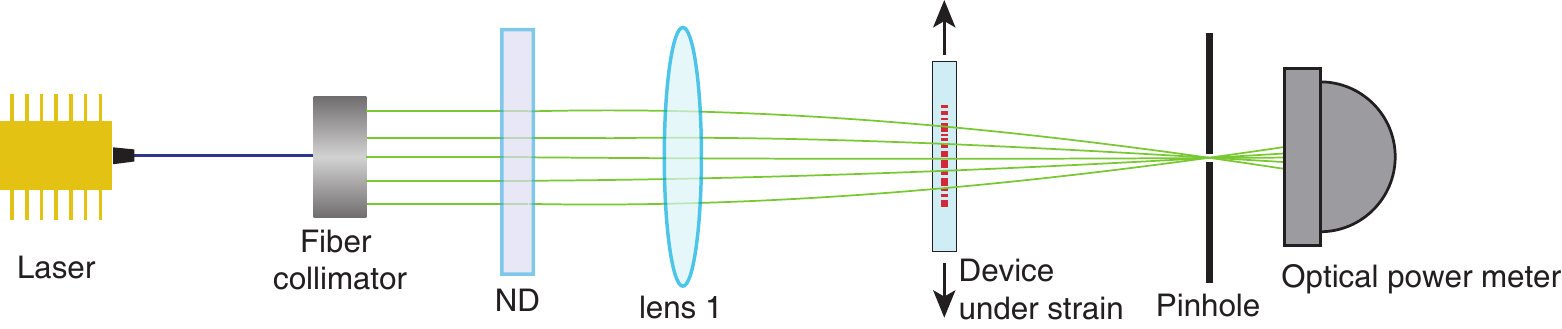}
\caption{Schematic illustration of the measurement setup used for measuring the efficiencies of the tunable microlens. ND: neutral density filter.}
\label{fig:S5_Setup}
\end{figure*}

\end{document}